# The Mass, Orbit and Location of Planet Nine Derived from Classical Astrophysics

## Abstract

In 2016 astronomers from the California Institute of Technology proposed the existence of a large, unseen 'Planet Nine' in a long period, eccentric orbit based on the analysis of the orbital pattern of 6 particular Trans Neptunian Objects (TNOs). Although they were not the first to come up with the idea, which goes back to the mid 1800s, they were the ones who found the best material evidence for it up to that point, in fact even better than they knew at the time. Using a computer simulation, the astronomers developed an estimate of Planet Nine's orbit, which they and others have analyzed and reanalyzed to determine likely celestial regions for optical and infrared search regions. To date, this approach has not located the planet, although an international team analyzing over 20 years of IRAS and AKARI satellite data of some intriguing indications of a large object moving through the constellation Taurus, the study published in May 2025. Taurus is the same region that this paper predicts Planet Nine to be in currently.

For the approach here, we use standard astrophysics to develop math models for estimating the mass, orbit and near term location of Planet Nine. Our models are based on the planet's presumed influence on the orbits of 12 TNOs: the original 6 analyzed by the CalTech astronomers and an additional 6 selected by the authors using similar although broader criteria. An additional goal of our approach is to make the rationale behind each claim regarding Planet X mathematically visible and verifiable to readers, who can use the models for further analysis with their own TNO data sets should they choose, which is rarely an option with simulations. Asteroid data used in our analysis came from the Jet Propulsion Lab's ephemeris generator and Johnston's Archive.

Using the math models described herein we found that each orbit of the 12 asteroids lies close to the same plane, presumably the planet's orbital plane. This geometric near-symmetry was found to uniquely determine the planet's orbital elements longitude of the ascending node at **+107.7°** and inclination of **+19.5°**. In azimuth the twelve asteroids' longitudes of perihelion were found to point to two stable clusters with their major axes forming a distinctive V pattern, whose bisecting center line points towards the planet's estimated aphelion. This pattern together with the cluster's center line was found to uniquely determine the planet's argument of perihelion at approximately **307.5°**. The planet's off axis Lagrange points L4 and L5 were found to play a key role in producing this clustering effect via an orbital process discovered by Romanian astronomers over 30 years ago. At least 4 and possibly 5 of the 12 asteroids appear to be in mean motion resonance with the planet, which led to an estimate of the orbit's semi major axis length of a **510 AU** and a period of **11512 years**.

Using a well-known equation for the region where orbiting bodies spend most of their time in orbit on average, the eccentricity for the planet was found to be approximately **0.39** implying a perihelion of **311 AU**, a wide orbital ellipse that completely surrounds half the asteroids' orbits nearest to their perihelia. And knowing the eccentricity, a quadratic expression for estimating the planet's mass was developed using a well known corollary to the law of conservation of angular momentum, which produced an estimate of **7.1 Earth masses**. The estimation for the planet's location also used the above mentioned corollary to develop estimates of the right ascension and declination of the planet's recent location, perihelion and aphelion, the first two locations estimated to be in the lower region of Taurus, the latter in the upper region of Libra.

On the whole, we discuss the quantifiable indications that something massive is arranging the orbits of 12 asteroids in our solar system in a measurably organized pattern. So many aspects of ordered behavior may well not be a coincidence.

Dr. Mattia A. Galiazzo mattia.galiazzo@gmail.com
Mr. Robert L. Finch rlfinch99@gmail.com 607-754-2630

[1] Independent researcher, Wien, Austria
[2] Senior Engineer ( Retired ), Lockheed Martin Corporation, Owego, New York, USA

## 1. Introduction

“Planet Nine”, originally designated “Planet X”, is a hypothetical planet believed for over two centuries to be orbiting the Sun somewhere in the outer reaches of space well outside the orbit of Pluto. In the 1980s an extensive effort to locate the proposed planet was undertaken by the Naval Observatory under the direction of Dr Robert Harrington (Harrington 1988), who used the Pioneer probes 10 and 11 to help with the search. Interest in the subject stalled following his death in 1993, but returned the next decade with a number of articles exploring the concept: (Collander-Brown, et al, 2000 ), (de la Fuente Marcos, 2014), ( Galiazzo, et al, 2014). An article in *Nature* appeared in 2014 by Gemini Observatory astronomers Chad Trujillo and Scott Sheppard who announced their discovery of the long-period dwarf planet 2012 VP113 (Trujillo & Sheppard 2014). That body provided additional evidence that the discoveries of two large mass, long period objects, the first being the dwarf planet Sedna discovered in 2003, might be indications of another much larger planetary body orbiting in the outermost reaches of the solar system. In February 2016 astronomers Michael Brown and Konstantin Batygin at the California Institute of Technology proposed the existence of a candidate for the distant planet, based on a computer simulation that analyzed a particular group of 6 Trans-Neptunian Objects (NTOs) orbiting in an unusually ordered pattern. Their goal was to explain the orbital configuration of the TNOs and to locate approximate regions where the hypothetical planet responsible might be observed (Batygin & Brown 2016, 2019).

For this analysis, rather than use simulations to analyze the existence of the unseen planet, we use mathematical and classical physics for our approach, and a larger data set of 12 Trans Neptunian Objects (TNOs), the original 6 from the CalTech study and an additional 6 selected by the authors using similar although broader orbital criteria. Finally, we developed additional models for predicting the planet’s mass and celestial coordinates of its recent location in Taurus along with the locations of its perihelion and aphelion, all using data relating to the author’s selected ‘reference day’ of September 1, 2020.

The orbit for Planet X developed here is very different from that of our earlier paper (Finch & Galiazzo 2020), due to our more recent discovery of the role played by the planet’s Lagrange points, which in particular allowed a considerably more refined calculation of the planet’s eccentricity. This effect is fully described below.

## 2. Methods

The fundamental assumption of the article is that the mass and orbital characteristics of the 12 reference asteroids analyzed provide sufficient information to approximate those of the unseen Planet X analytically. The fidelity of the estimates is limited only by the quality of the reference data and the number of asteroids used in the analysis.

The mathematical models described herein were developed using basic astrophysical principles, reference asteroid data from the Jet Propulsion Lab’s ephemeris generator, Johnston’s Archive and standard gravitational dynamics. Of the 12 reference TNO’s analyzed, the first 6 in Table 1 are those analyzed by Brown and Batygin in their original article; the remaining 6 have been selected by the authors using similar although broader criteria than those associated with the original CalTech data set to include several additional TNO’s whose orbits are also being affected by Planet X but that the CalTech astronomers had excluded. All asteroid data used in our analysis is associated with the ‘reference day’ of September 1, 2020. In the CalTech data set all asteroids had orbits that fall within a spread of 380° to 480° in longitude of perihelion, inclinations above 0° and below 30°, semi major axis lengths lying between 300 AU and 510 AU, and orbital periods of at least 3400 years. For our selection process which allowed inclusion of the additional 6 TNO’s being affected, the spread in longitude of perihelion was found to be between 350° and 480°, inclinations between 0° and 35°, semi major axis lengths lying between 150 AU and 540 AU and orbital periods of 1800 years and greater. There were no restrictions in either data set placed on asteroids’ mass or eccentricity.

The mass of most of the TNOs used in all Planet X studies is not currently known accurately. For the analysis described herein, the authors relied on the currently accepted diameter estimates of each body, together with density approximations of TNOs and centaurs with similar size, mass and density listed in Johnston’s Archive (Johnston 2019). See (Finch and Galiazzo 2020) for details on the method. For reference asteroids with diameters greater than 900 km, the densities were estimated to be 1.96 g/cm$^3$; for diameters less than 900 km but greater than or equal to 400 km, the associated estimate was 1.06 g/cm$^3$; and for diameters less than 400 km the estimate was 0.94 g/cm$^3$. A published mass window is available for the dwarf planet Sedna, from which we used the more conservative, smaller value of the two limits at $1.70 \times 10^{21}$ kg. For the mass of the second largest body of the 12 in our analysis, the dwarf planet 2012 VP113, we used its published diameter together with the same density as that implied by the above mass value and diameter of Sedna, which yields a mass approximation for the second planetoid of $0.43 \times 10^{21}$ kg. In calculating an approximate diameter of Planet X, the density of the Earth was presumed to provide a lower bound, since Earth has the highest density of the known planets.

## 2.1 The Orbital Elements of Planet X

The foundation of the model rests on the geometric relationship of the asteroids' orbits to each other and to the orbital elements of the hypothetical Planet X. Central to estimating both is the transformation matrix $\boldsymbol{U}$ associated with each body, formed from the body's standard orbital elements: inclination ($\iota$), longitude of the ascending node ($\Omega$) and argument of perihelion ($\omega$). The $\boldsymbol{U}$ matrix transforms a Cartesian 3x1 vector (e.g. $\hat{I}, \hat{J}, \hat{K}$) in the heliocentric equatorial coordinate system into the orbiting body's local coordinate system (e.g. $\hat{x}, \hat{y}, \hat{z}$). The inverse of $\boldsymbol{U}$, which is simply its transpose $\boldsymbol{U}^T$, performs the reverse transformation.

$$\boldsymbol{U} = \begin{matrix} \cos\Omega\cos\omega - \sin\Omega\cos\iota\sin\omega & \sin\Omega\cos\omega + \cos\Omega\cos\iota\sin\omega & \sin\iota\sin\omega \\ -\cos\Omega\sin\omega - \sin\Omega\cos\iota\cos\omega & -\sin\Omega\sin\omega + \cos\Omega\cos\iota\cos\omega & \sin\iota\cos\omega \\ \sin\iota\sin\Omega & -\sin\iota\cos\Omega & \cos\iota \end{matrix} \quad (1)$$

The axes of the equatorial coordinate system are oriented such that the components $\hat{I}$ and $\hat{J}$ lie in the equatorial plane of the central body with $\hat{I}$ pointing in the direction of the vernal equinox and $\hat{J}$ perpendicular to $\hat{I}$. These two axes define the planet's reference plane of the ecliptic. The third axis $\hat{K}$ is perpendicular to the reference plane. For the orbital plane, $\hat{x}$ and $\hat{y}$ lie in the plane with $\hat{x}$ pointing in the direction of perihelion, the $\hat{z}$ axis being perpendicular to the plane of the orbit and $\hat{y}$ being perpendicular to both $\hat{x}$ and $\hat{z}$.

For the following analysis the $\boldsymbol{U}$ matrix describes the geometric relationship associating a body's orbital plane with the reference equatorial coordinate system – effectively the body's orbital orientation in space. Each body in the following analysis has its own $\boldsymbol{U}$ matrix, and two rows of each are special. The first row of $\boldsymbol{U}_p$ is the unit vector $\boldsymbol{w}_p$ which points along the orbit's major axis towards perihelion; the third row is the unit vector $\boldsymbol{u}_p$ perpendicular to the body's orbital plane and parallel to its angular momentum vector; the second row vector is normal to both $\boldsymbol{u}_p$ and $\boldsymbol{w}_p$.

The orbital elements of the hypothetical Planet X were determined directly from the components of its $\boldsymbol{U}$ matrix. In the coordinate system of the orbit, the unit vector normal to the orbital plane is simply $(0, 0, 1)^T$. When this vector is transformed to the heliocentric equatorial system using $\boldsymbol{U}_p^{\,T}$ the resulting vector is perpendicular to the body's orbital plane and is equal to the third row of the $\boldsymbol{U}_p$ matrix, a key vector used throughout the analysis that we refer to from here on as the "normal unit vector" $\boldsymbol{u}$ with an appropriate subscript. As will be shown with the data, the orbital planes of all the reference asteroids are very similar in orientation, indicating that Planet X is influencing the reference asteroids in a very orderly way. This ordering is most obvious regarding the alignment of the asteroids' orbital planes, presumably with that of the hypothetical Planet X. Accordingly, if we let $\boldsymbol{u}_p$ represent the normal unit vector of Planet X perpendicular to its plane, and we assume the associated unit vectors of the reference asteroids to be normally distributed, a standard estimate of $\boldsymbol{u}_p$ is simply the average of the normal unit vectors $\boldsymbol{u}_i$ of the reference asteroids normalized to a unit vector given in equation (2) below. By knowing $\boldsymbol{u}_p$ the planet's longitude of perihelion and inclination are completely determined and can be calculated using the following steps leading to expressions (4) and (5):

$$\boldsymbol{u}_p = \Sigma\,\boldsymbol{u}_i \,/\, \|\,\Sigma\,\boldsymbol{u}_i\,\| \quad (2)$$

$$= \sin\iota_p\sin\Omega_p \quad -\sin\iota_p\cos\Omega_p \quad \cos\iota_p \quad (3)$$

$$\tan\Omega_p = \boldsymbol{u}_p(x) \,/\, -\boldsymbol{u}_p(y)\,, \quad (4)$$

$$\tan\iota_p = (\boldsymbol{u}_p(x)^2 + \boldsymbol{u}_p(y)^2)^{0.5} \,/\, \boldsymbol{u}_p(z)\,. \quad (5)$$

The angle $\zeta_i$ between the orbital planes of Planet X and a given reference asteroid $i$ is equal to the angle between the normal unit vectors of the two bodies, and can be computed by taking the scalar product of the two associated vectors. The trigonometric form (7) below is then derived directly from the scalar product of the associated algebraic components of the third row of the $\boldsymbol{U}$ matrices of Planet X and asteroid $i$, described generally in the matrix (1):

$$\cos\zeta_i = \boldsymbol{u}_p \bullet \boldsymbol{u}_i \quad (6)$$

$$= \sin\iota_p\sin\iota_i\cos(\Omega_p - \Omega_i) + \cos\iota_p\cos\iota_i\,. \quad (7)$$

Using this equation it was found that the average deviation angle $\zeta_i$ of the asteroids' orbital planes from that of Planet X for the 12 reference asteroids is only about 13.5° with a maximum of about 20.0°. The planes of the reference asteroids are clearly highly correlated. The inclination and longitude of the ascending node for the planet were found using (4) and (5) above to be +14.6 ° and +107.7 °, respectively, using all 12 reference asteroids, which specifies the orbital plane uniquely. This planar orientation, and the analogous estimate using the original 6 asteroids (see Table 1), will be presumed throughout the discussion.

To estimate the argument of perihelion we need to have an idea about the azimuth orientation of the major axis of the planet's orbit. The dominant effect of Planet X on the reference asteroids appears to be to align their orbital planes with that of Planet X, but another effect described in more detail below has caused a shift in each asteroid's azimuth orientation as well. Smaller though that influence may be at great distances, the data clearly indicates that the effect has caused two clusters to develop with the azimuth orientations in the reference asteroids' orbits. This dual effect is clearly evident when we examine the longitudes of perihelion $\varpi_i$ of the 12 asteroids provided in column 11 of Table 1 below. One cluster occurs in the interval of 355° to 400°, the other between 430° and 480°, both clusters being close to the same angular 'width'. Note that the values of $\varpi_i$ are expressed in angular ranges of 180° to 500° throughout to make the angular clustering and its relationships more apparent.

The azimuth orientation of an asteroid's orbit is a function of all three orbital coordinates but for small inclinations is predominantly due to the argument of perihelion $\omega_i$ and the longitude of the ascending node $\Omega_i$. This fact is readily apparent by examining the first row of the $\boldsymbol{U}$ matrix, referred to below as the azimuth unit vector $\boldsymbol{w}$, which points along an orbit's central axis towards perihelion. Note that when the body's inclination is small the first two components of $\boldsymbol{U}$'s first row are approximately the cosine and sine of the longitude of perihelion, respectively, both perturbed slightly by the cosine of inclination, which is relatively close to unity for all reference asteroids being considered. The result is that gravitational forces of each asteroid's azimuth orientation are shifted toward aligning with one or the other of the clusters' two dominant orbital directions, whichever has the smallest angular separation from the asteroid's major axis. Planet X's azimuth orientation is therefore most likely to be aligned mid way between the two clusters, which we describe in more detail below.

Presuming this to be true, the associated unit vector $\boldsymbol{w}_p$ describing this direction is relative easy to estimate. To find the azimuth vector that approximately divides the two clusters 'evenly', we have to find the argument of perihelion that accomplishes that when using the planet's inclination and longitude of the ascending node estimates from (4) and (5) above. Equations (8) and (9) below estimate the average angular deviation $\Delta_p$ from the probable center vector of each of the two clusters from the planet's major axis, where the two weighted summations represent the averages of the larger ($l$) and smaller ($s$) values of the longitude of perihelion of each cluster, respectively. From the approximation of $\Delta_p$ the argument of perihelion in (10) follows directly from the following sequence:

$$(1/m)\,[\Sigma_s(\Omega_s + \omega_s)] + \Delta_p = (1/n)\,[\Sigma_l(\Omega_l + \omega_l)] - \Delta_p, \tag{8}$$

$$\Delta_p = 0.5\,[(1/n)\,\Sigma_l(\Omega_l + \omega_l) - (1/m)\,\Sigma_s(\Omega_s + \omega_s)], \tag{9}$$

which allows estimating the planet's argument of perihelion as

$$\omega_p = (1/n)\,\Sigma_l(\Omega_l + \omega_l) - \Omega_p - \Delta_p. \tag{10}$$

Note that when the longitude of the ascending node and azimuth orientation are known, as they are from the above development, the argument of perihelion is completely determined. Given this fact the argument of perihelion was estimated to be 307.3° using all 12 asteroids.

Adding $\Omega_p$ to both sides of (10), the planet's longitude of perihelion at the center point of the two clusters follows accordingly from two of the planet's orbital elements:

$$\varpi_p = \Omega_p + \omega_p. \tag{11}$$

Using the three estimated orbital parameters $\iota_p$, $\Omega_p$ and $\omega_p$, the associated $\boldsymbol{U}$ matrix for Planet X can now be formed using equation (1). And with this result the planet's formal azimuth unit vector $\boldsymbol{w}_p$ follows from (1), being the first row of the associated $\boldsymbol{U}$ matrix, which lies along the orbit's main axis pointing towards perihelion:

$$\boldsymbol{w}_p = \cos\Omega_p\cos\omega_p - \sin\Omega_p\cos\iota_p\sin\omega_p \quad \sin\Omega_p\cos\omega_p + \cos\Omega_p\cos\iota_p\sin\omega_p \quad \sin\iota_p\sin\omega_p. \tag{12}$$

With the above formalism, the longitude of perihelion of the planet was estimated to be 428.7° using the original 6 TNOs and 415.1° using all 12. The orbital component estimates using both data sets are given in Section 4.0.

## 2.3 The Orbital Eccentricity, Semi Major Axis and Period

Astronomy has two ways of determining the average range of the planet from its major focus over its orbit, depending on which characteristic is most relevant to the problem at hand. For our purposes the most useful approach is the average of the planet's range taken over time, the region in which the planet spends the greatest part of its time per unit arc length in orbit on average. For the associated range of Planet X we use the well known expression for this time-averaged range as follows:

$$\bar{R}_p = a_p \left( 1 + \tfrac{1}{2} e_p^{\,2} \right). \tag{13}$$

The two points on the planet's orbit having this value of range correspond to two orbital regions in its inbound and outbound paths, regions where the planet's range changes relatively slowly. The time-averaged range occurs when the planet is between aphelion and the two points of the minor axis that intersect the planet's orbit, at which point both the angular rate of the range and orbital velocity are changing at a relatively low rate. When we know the angular orientation from the Sun to any orbital location, the associated range values can be derived from the Kepler equation for the body's range from the primary focus to the body's true anomaly, $\theta_p$ being measured counter clockwise from perihelion to the range vector $R_p$ :

$$R_p = a_p \left( 1 - e_p^{\,2} \right) / \left( 1 + e_p \cos \theta_p \right). \tag{14}$$

By equating expressions (13) and (14), we can easily derive an expression for the planet's true anomaly at this point in its orbit in terms of its eccentricity, independent of its semi major axis length, unknown to us at this stage in the development:

$$\cos \theta_p = 3\, e_p / \left( 2 + e_p^{\,2} \right). \tag{15}$$

To estimate the planet's eccentricity $e_p$ therefore, we first need to estimate its true anomaly at the point in its orbit associated with its time-averaged range. Because the asteroids' average range occurs in the same two regions of space on either side of the planet's major axis, the planet's time averaged range should occur in approximately the same region, which we presume for the following estimation process. Note that all asteroid ranges are presumed to occur in a single grouping to derive maximum benefit of the limited number of data points available, which is justified due to the symmetry of the asteroids' two clusters and by the fact that the cosine of the true anomalies symmetrically positioned on either side of the planet's major axis are equal in both magnitude and sign. The planet's true anomaly is therefore taken as that angle associated with the average of the asteroids' time-averaged range values plus the angular difference between each asteroid's major axis and that of Planet X:

$$\theta_p = \Sigma \left[ \, |\Delta_i| + (1/n)\, \Sigma \arccos \left[ 3\, e_i / \left( 2 + e_i^{\,2} \right) \right] \right], \tag{16}$$

where n is the number of asteroids being used in the calculation. The planet's eccentricity $e_p$ *i*s then computed by inverting equation (15) with a truncated series approximation to arrive at the following estimate:

$$e_p = (2/3) \cos \theta_p + (1/6) \cos^3 \theta_p \,. \tag{17}$$

The average eccentricity of the 12 asteroids is 0.83, so we would expect the estimate of Planet X's eccentricity to be considerably smaller, which it is. Using (17) the estimates are then 0.37 using the original 6 asteroids and 0.39 using all 12.

So, what is it that causes this double clustering to occur anyway? To provide an idea about this, a well established fact regarding circular and nearly circular planetary orbits is that the planet's gravitational attractions can pull asteroids into its orbit, which over time form two clusters around the planet's off axis Lagrange points L4 and L5. Normally, a planet and the Sun exert an unbalanced gravitational force at a point, altering the orbit of whatever is near the point, but at the Lagrange points the gravitational force of these two bodies and the associated centrifugal force are balanced. With this dynamic these bodies produce five such Lagrange points, all lying in the plane of the orbit with three on the axis between the bodies (L1, L2 and L3) and the other two located off axis on the planet's orbit, which remain a fixed distant apart and equidistant from the planet and the system's barycenter. These two Lagrange points pull nearby asteroids into the planet's orbit, where they form two symmetric clusters called 'Trojan asteroids', one cluster orbiting ahead of the planet's motion, the other trailing behind. The symmetry is most clearly observed by two contiguous equilateral triangles formed by four vertices made up of the orbiting planet, the two off axis Lagrange points and the barycenter, all of which rotate along with the planet's orbit. The planets in our solar system have a large number of such Trojans asteroids; Jupiter in particular has several thousand in each of its two clusters, and even the Earth has two at current count, both at the Lagrange point L4.

For planets having pronounced elliptical orbits, a related process occurs but in a substantially different manner. The off axis Lagrange points are still equidistant from the planet and the barycenter, but no longer remain a fixed distance from the orbit's major focal point but rather move along the orbit towards the Sun, coming closer together as the planet approaches perihelion. The two triangles are no long equilateral but isosceles, with variable degrees of acuteness during the orbit as the Lagrange points help to form the V-shaped pattern of the reference asteroids' major axes. Near perihelion the asteroids orbit well inside the planet's orbit, pulling the asteroids from the two clusters toward the planet, which causes their orbital axes to revolve outwardly about the primary focus as the planet approaches and exits perihelion. For most of its orbit the planet remains 'outside' the asteroids' orbits, but as it approaches aphelion the planet crosses into the inner part of the V pulling the asteroids' orbits inwardly toward the planet's major axis, canceling the outward revolution occurring near perihelion and so stabilizing the V-shaped pattern over time. Technical details of this process were described in the early 90s in a paper by Romanian astronomers (Todoran & Roman 1992).

Another interesting fact about this process is that the centripetal acceleration of the reference asteroids has also become "ordered". If we examine these accelerations at the points of their smallest value at their minor axes, at which point the range is equal to the semi major axis length, we see that the acceleration values are steadily larger the closer these axes are to perihelion and vice versa. The centripetal acceleration at this point in the orbit is given in the first parentheses in (19) by the square of the body's instantaneous velocity $v_i$ from the classical vis viva equation, $R$ being the body's instantaneous range and $a$ its semi major axis length:

$$v^2 = GM\,(2/R - 1/a)\,. \tag{18}$$

On setting the range in (18) to the semi major axis length as indicated above and then dividing by the radius of curvature $R_{ci}$ at this point, we have the following:

$$v_i^2 / R_{ci} = (GM / a_i) / (a_i / (1 - e_i^2)^{0.5}) \tag{19}$$

$$= GM\,(1 - e_i^2)^{0.5} / a_i^2 \tag{20}$$

where $G$ is the gravitational constant, $M$ is the mass of the Sun, $a_i$ is the asteroid's semi major axis length and $e_i$ is its eccentricity. The centripetal acceleration of Planet X at the minor axis is obviously within the range of that of the asteroids. To find the planet's semi major axis length we use the average of the square root of the asteroids' inverse accelerations, scaled by its eccentricity term as indicated in (20) above :

$$a_p = (1 - e_p^2)^{0.25}\,(1/n) \sum a_i / (1 - e_i^2)^{0.25}. \tag{21}$$

Using this expression the estimation produced semi major axis length of 519 AU for the case of the 6 asteroids and 450 AU for the case with 12. Applying Kepler's third law, the orbital period associated with the semi major axis length 519 AU is 11820 years. For 12 asteroids, the corresponding estimate of 450 AU similarly produces an orbital period of 9550 years. The reason why the two numbers are so different is because the data set of 12 has three asteroids with semi major axes lengths less than 200 AU, and CalTech's original data set has none. As mentioned earlier the data sets are simply not comprehensive enough to allow an estimate of higher fidelity. With this approach we can say that the axis length is in the vicinity of 440 AU to 540 AU. That said there is an interesting way to improve this uncertainty.

Examining the periods of the reference asteroids, there are four asteroids that are near being in mean motion resonance with Planet X based on the above estimates: Sedna, 2000 $OO_{67}$, 2013$RF_{98}$, and 2004 $VN_{112}$, the first two being in a 1:1 resonance the second two in a 1:2 resonance. These apparent resonances are most likely of higher fidelity than that associated with the centripetal accelerations, so we can use this indication to improve on those estimates. If we then take the average of the weighted resonance periods of these four, the period estimate and the likely resonances are more pronounced:

$$T_p = \tfrac{1}{4}\,(11390 + 11760 + 2\,(5600) + 2(5856)) = 11512 \text{ yrs.}$$

Presuming this is indeed the period of Planet X, all resonance periods are within 2% of this value, and the semi major axis length is 510 AU as derived from Kepler's third law. And so the planet has apparently created yet another synchronistic pattern among the reference asteroids. Other possible resonances, neither of which were used in the above estimate, involve that of the dwarf planet 2012 VP113, which appears to be in a 2:5 resonance with Planet X as well, and the more recently announced discovery of dwarf planet 2017 OF201, which may be in a 2:1 resonance with Planet X.

Continuing with the case of the 12 reference asteroids, while knowing the eccentricity and the semi major axis length ,we have the planet's perihelion $a_p\,(1 - e_p)$, which gives the value 311 AU.

## 2.4 System Mass Distribution and Mass of Planet X

When an isolated system of bodies is moving solely under its own mutual gravitational attraction with no other influences, the system's angular momentum is obviously conserved. Less well known is that its collective mass distribution maintains a certain symmetry as well. Whenever we multiply the vector distance of each body, as measured from the coordinate origin at the system's center of mass (CM), by the body's mass the sum of the collection of mass-weighted vectors remains zero. (For the following analysis, distances to the 'right' of the CM are defined as positive, to the 'left' as negative.) Because the Sun has over 99.9% of the mass of the inner solar system and the approximation of Planet X's orbit implies that the planet's trajectory extends far outside the inner planets at its closest approach, our model of the solar system treats the inner solar system within Pluto's orbit as a single large mass centered at the Sun. By assumption, Planet X and the set of reference asteroids are the only remaining bodies the model then has to take into account. The mass of Planet X orbiting around the Sun creates a reciprocal rotation of the Sun, and in the absence of any other objects the two bodies orbit about their common center of mass, which orbits the Sun as well. The reference asteroids are shown to have a collective mass several orders of magnitude smaller than that of Planet X, and even though their contribution to the overall mass distribution principle is extremely small, the system's center of mass is shifted ever so slightly nevertheless, which also shifts the orbit of the two body Sun- Planet X system. This slight shift is what allows the mass and location of Planet X to be estimated directly. We provide this analysis with more detail since to our knowledge no Planet X article has included a credible methodology for estimating the planet's mass.

Given this preamble the arrangement of the system can now be described using the heliocentric equatorial coordinate system with moving origin at the system's center of mass. Let $\boldsymbol{R}_p$ , $\boldsymbol{R}_M$ and $\boldsymbol{R}_i$ represent the range vectors in that coordinate system from the system's center of mass, the origin of the above coordinate system, to Planet X, the Sun and each reference asteroid, respectively, and let $m_p$, $M$ and $m_i$ represent the mass of Planet X, the Sun and each of the associated reference asteroids. For the following analysis, the range from the CM to Planet X is defined to be positive, and to the Sun and the reference asteroids to be negative. Given these conventions the basic model of the system of mass-weighted range vectors described above can be written with the following simple expression:

$$m_p \boldsymbol{R}_p \;+\; M \boldsymbol{R}_M \;+\; \Sigma\, m_i \boldsymbol{R}_i \;=\; \boldsymbol{0}. \qquad (22)$$

The first two terms of (22) comprise the vector difference of two large quantities nearly equal in magnitude and opposite in sign, which we wish to refine mathematically. If we begin by taking the scalar product of the entire expression by $\boldsymbol{R}_p$ , noting that $\boldsymbol{R}_p$ and $\boldsymbol{R}_M$ are effectively co-linear, and then dividing each term by the magnitude $R_p$ and rearranging a bit, we obtain the following:

$$m_p R_p \;-\; M R_M \;=\; \Sigma\, m_i R_i \cos \varphi_i \;, \qquad (23)$$

where $\varphi_i$ is the angle between the range vectors of Planet X and asteroid $i$. We begin the estimation with assuming no asteroids present, in which case we have the following sequence:

$$0 \;=\; m_p R_p \;-\; M R_M \qquad (24)$$

$$= (\; m_p R_p \;-\; M R_M \;)\; M \,/\, (\, M + m_p \,) \qquad (25)$$

$$= R_p\, M\, m_p / (\, M + m_p \,) \;-\; R_M\, M^2 \,/\, (\, M + m_p \,) \qquad (26)$$

$$= R_p\, M\, m_p / (\, M + m_p \,) \;-\; R_M\, M \;+\; R_M\, M\, m_p \,/\, (\, M + m_p \,). \qquad (27)$$

The third term of (27) is at least four orders of magnitude smaller than the other two, so we omit that one as negligible, and on subtracting the modified expression from (24) we have the following:

$$0 \;=\; m_p R_p \;-\; R_p\, M\, m_p / (\, M + m_p \,) \qquad (28)$$

When we now reintroduce the reference asteroids, we implicitly modify the value of $R_p$ to preserve the stated equality accordingly and proceed with the following steps:

$$m_p R_p \;-\; R_p\, M\, m_p / (\, M + m_p \,) \;=\; \Sigma\, m_i R_i \cos \varphi_i \qquad (29)$$

$$(\, M + m_p \,)\; [\; m_p R_p \;-\; R_p\, M\, m_p / (\, M + m_p \,)\; ] \;=\; \Sigma\, m_i R_i \cos \varphi_i \qquad (30)$$

$$m_p^2 R_p = (M + m_p) \Sigma m_i R_i \cos \varphi_i \quad (31)$$

Again the second implied term involving $m_p$ on the right is several orders of magnitude smaller than the other, so we omit that one as negligible also, arriving finally at the following expression:

$$m_p^2 R_p = M \Sigma m_i R_i \cos \varphi_i \quad (32)$$

with the vector form to be used later following accordingly:

$$m_p^2 \boldsymbol{R}_p = -M \Sigma m_i \boldsymbol{R}_i . \quad (33)$$

Equations (32) and (33) are valid for all locations of the asteroids and Planet X over the orbits of these bodies. It has other uses as we show below, but to use it for estimating the mass of Planet X we have to know more. In particular since the equation is applicable at all locations of the bodies involved it must also be applicable at their average locations as well. To calculate this average we use the classical expected value of the range of each body's orbital location position $R$, with averages taken over the true anomaly, well known to be the length of the orbit's semi minor axis. Because the average range occurs twice in each orbit, we assume that its position is along the orbit's major axis. Proceeding accordingly, we have the desired estimate of Planet X's mass:

$$b_i = a_i (1 - e_i^2)^{0.5} \quad \text{for asteroid } i \text{, and for Planet X} \quad b_p = a_p (1 - e_p^2)^{0.5}$$

$$m_p^2 = M (\Sigma m_i b_i \cos \gamma_i) / b_p , \quad (34)$$

where $\gamma_i$ is the angle between the major axes of Planet X and asteroid $i$. The estimate of mass increases as the square root of the collective mass of all asteroids of the under the planet's organizing influence, presumed here to be only the reference asteroids used in the study. Inasmuch as the total number and mass of asteroids under the planet's similar influence is not completely known, the fidelity of the estimate clearly improves as the number of such asteroids used in the estimation process increases. As the final result, using (34) we estimated the planet's mass to be about 7.4 Earth masses using the original 6 asteroids and 7.1 using all 12.

## 2.5 Location of Planet X

The location of celestial bodies are typically described using celestial coordinates defined relative to the Earth's equatorial plane by the right ascension and declination. The ephemeris for each reference asteroid is available from the Jet Propulsion Laboratory's web site, whose ephemeris generator was used to obtain the reference asteroids' celestial positions for our reference date September 1, 2020, chosen as noted above as the reference date for the analysis. Knowing these coordinates for a celestial object allows computing its coordinates in the inertial rectangular equatorial coordinate frame used for the following analysis. Accordingly, if we let $\alpha_i$ be the right ascension of a particular reference asteroid, $\delta_i$ its declination, and $\boldsymbol{R}_i$ its vector distance from the Sun with magnitude $R_i$, then the corresponding rectangular equatorial coordinates are computed as follows:

$$\boldsymbol{R}_i = \begin{matrix} R_i \cos \delta_i \cos \alpha_i \\ R_i \cos \delta_i \sin \alpha_i \\ R_i \sin \delta_i \end{matrix} \quad (35)$$

To determine the location of Planet X we first need to compute $\boldsymbol{R}_i$ using equation (35) for each of the 12 reference asteroids corresponding to the reference day and using epoch (2000), the celestial coordinates corresponding to these conditions provided in Table 1. This collection of range vectors are then used to estimate the location vector $\boldsymbol{R}_p$ of Planet X for the reference day below, from which the planet's right ascension and declination are derived as shown. The planet's position vector is first approximated using the planet's mass estimate and the mass-weighted range vectors of the reference asteroids. After rearranging equation (33) slightly, the current position of Planet X is then estimated as

$$\boldsymbol{R}_p = (M / m_p^2) \Sigma m_i \boldsymbol{R}_i , \quad (36)$$

where $\boldsymbol{R}_p$ and $\boldsymbol{R}_i$ are the position vectors of Planet X and asteroid $i$ at the reference date in 2020. The associated location unit vector $\boldsymbol{l}_p$ in equatorial coordinates is then calculated using the above location equation which gives the following:

$$\boldsymbol{l}_p = \boldsymbol{R}_p / R_p \quad (37)$$

where $R_p$ is the magnitude of $\boldsymbol{R}_p$. The planet's right ascension and declination are then estimated from the unit vector's components directly, noting that whenever $w_x < 0$, 180° has to be added to $\alpha_p$ to obtain the correct quadrant:

$$\alpha_p \;=\; \tan^{-1}\; l_{py} \;/\; l_{px}\,, \tag{38}$$

$$\delta_p \;=\; \tan^{-1}\; l_{pz} \;/\; (\; l_{px}^{\;2} \;+\; l_{py}^{\;2}\;)^{\,0.5}\,. \tag{39}$$

As we discuss in Section 3.0 there are data limitations in using this approach for locating Planet X, but we believe we may have that solved for our given geometry. As a result we estimate the planet to have been at right ascension 3h 52m 41.0s and declination +7° 22' 48.0" on our reference day, which places the planet's celestial coordinates on that day in the lower region of Taurus (Fig.1).

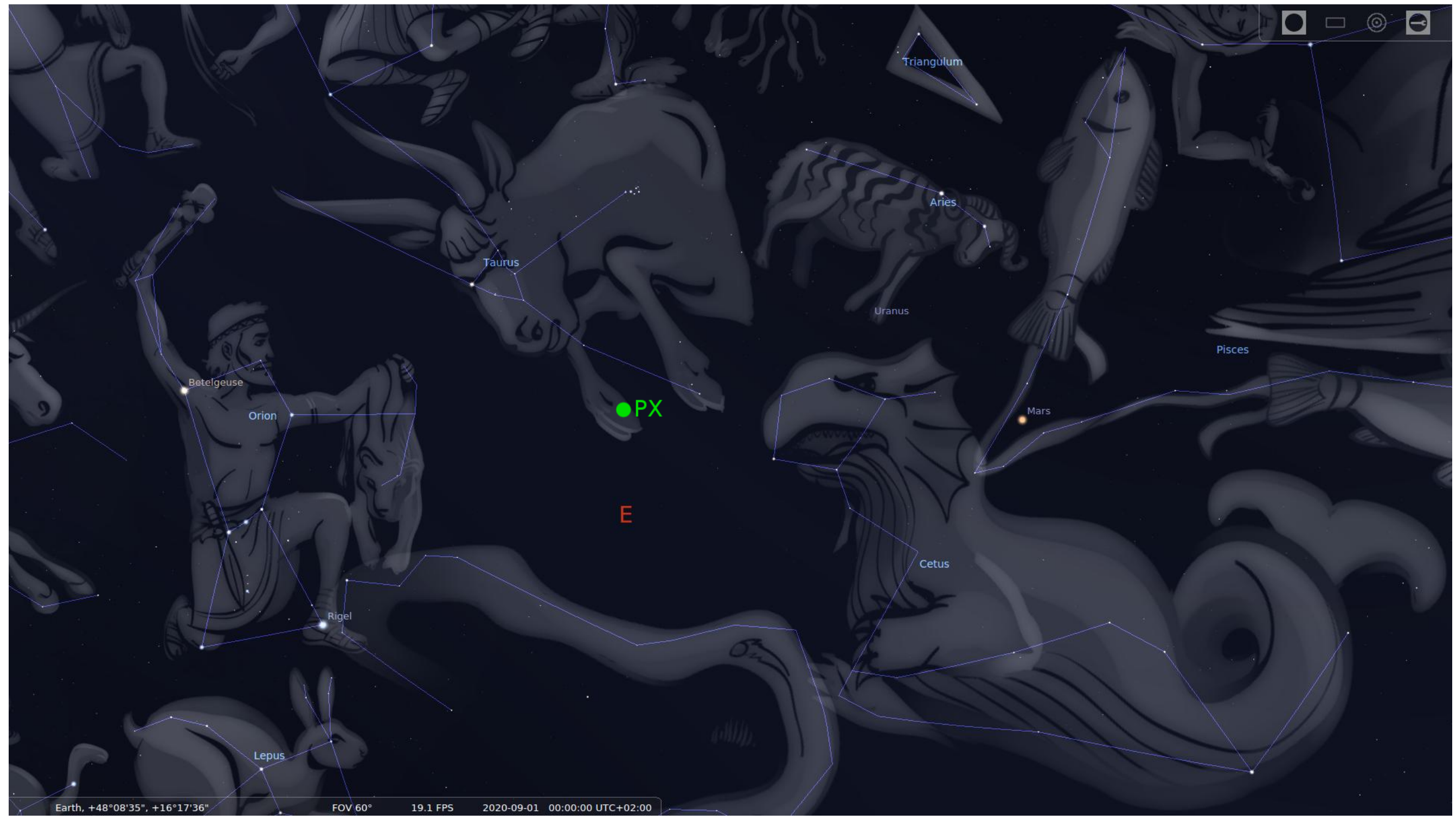

**Figure 1** The green point shows the approximate position of PX at Sept. 1 2020 00:00 UT in the constellation of Taurus.

At theThat time, one of the closest stars was X-Tau; see Figure 2.

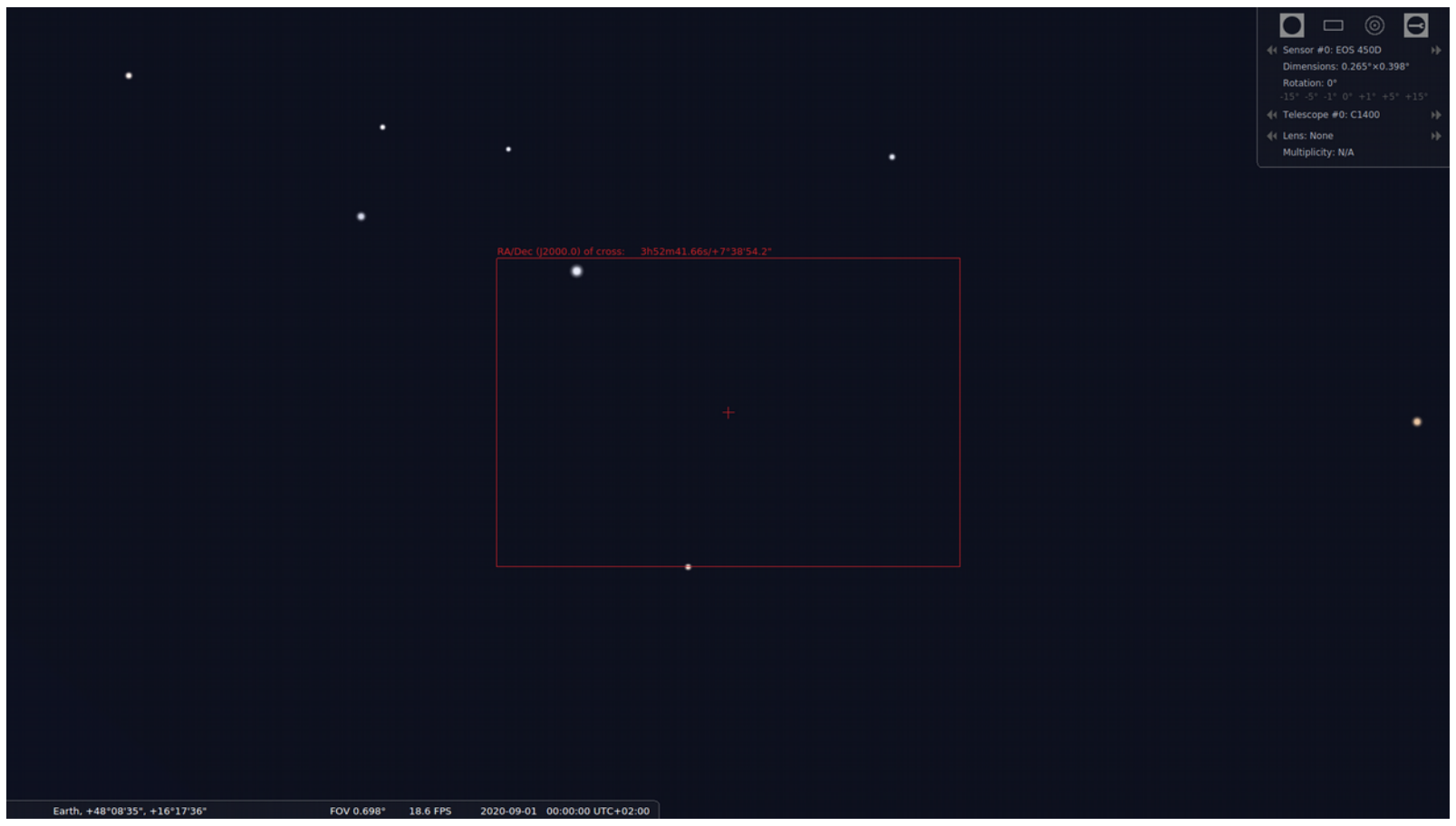

**Figure 2** The red crox shows the position of PX at September 1 2020 UT. One of the closest stars here is X-Tau (SAO111476 or HD24100): it is the one up right almost on the corner. (Done with Stellarium)

## 2.6 Perihelion, Aphelion Location and Mean Anomaly

To consider the estimated position coordinates in a broader context, the location of the planet's perihelion and aphelion are also useful to estimate for comparison. For this we first need to transform the unit azimuth vector $\boldsymbol{w}_p$ from the ecliptic coordinate system pointing along the orbit's major axis, given by the first row of the planet's ***U*** matrix although treated below as a column vector, into the equatorial coordinate system. This calculation uses the Earth's tilt angle ε, the obliquity of the ecliptic currently 23.44°, to rotate the unit vector by this amount as follows:

$$v_p = \begin{pmatrix} \cos\omega\cos\Omega - \sin\omega\sin\Omega\cos\iota \\ \cos\varepsilon(\cos\omega\sin\Omega + \sin\omega\cos\Omega\cos\iota) - \sin\varepsilon(\sin\omega\sin\iota) \\ \sin\varepsilon(\cos\omega\sin\Omega + \sin\omega\cos\Omega\cos\iota) + \cos\varepsilon(\sin\omega\sin\iota) \end{pmatrix} = \begin{pmatrix} 1 & 0 & 0 \\ 0 & \cos\varepsilon & -\sin\varepsilon \\ 0 & \sin\varepsilon & \cos\varepsilon \end{pmatrix} \begin{pmatrix} \cos\omega\cos\Omega - \sin\omega\sin\Omega\cos\iota \\ \cos\omega\sin\Omega + \cos\Omega\sin\omega\cos\iota \\ \sin\omega\sin\iota \end{pmatrix} \quad (40)$$

The unit vector $v_p$ is implicitly equal to the following form comprised of the celestial coordinates of particular locations on Planet X's orbit, which we wish to determine, in this case the perihelion and aphelion.

$$\boldsymbol{v}_p = \begin{pmatrix} \cos\delta_p \cos\alpha_p \\ \cos\delta_p \sin\alpha_p \\ \sin\delta_p \end{pmatrix} \quad (41)$$

The right ascension and declination of the planet's aphelion and perihelion are then calculated using the unit vector in (41). Note that the positions of these points in rectangular coordinates are equal in magnitude and opposite in sign, which have to be used for calculating the celestial position angles.

Right ascension: $\tan\alpha_p = v_y / v_x$ (42)

Declination: $\tan\delta_p = v_z / (v_y^2 + v_x^2)^{0.5}$ (43)

Again, whenever $v_x < 0$, the above computed value of α is increased by 180° to obtain the correct quadrant.

The perihelion's proximity to the planet's location on the reference day can be readily estimated. The angle $\beta$ between the vector pointing along the orbit's major axis and the range vector of Planet X on the reference day is calculated by taking the scalar product of these two vectors, described in equations (12) and (36) and normalizing the result to a unit vector. When the planet is near its perihelion, the associated distance $S_p$ along the planet's orbit between the two points is given by

$$S_p = p\ \beta. \quad (44)$$

From this equation and the standard Kepler orbital velocity equation (vis viva), the elapsed time $\Delta T$ between perihelion and the reference day's position can be easily estimated as follows:

$$\Delta T = (p\ \beta) / (GM(2/R_p - 1/a_p))^{0.5} \quad (45)$$

Given that in the planet's orbital period $T_p$ the planet sweeps out 2π radians, then the average rate of sweep $n$ is then

$$n = 2\pi / T_p. \quad (46)$$

## 3.0 Orbital Evolution Analysis

We simulated the orbits of PX and Sedna numerically with the 4 major planets included, once with PX present and again without PX, and we compared the orbital elements to verify that the system is still stable with PX present. We used the Lie-Integrator (Dvorak, Hanslmeier, 1984 ) for a time frame equal to 1.5 million years with an integration step size of 10 thousand years. The results of this simulation analysis are presented in Figures 3 through 6 below. Figure 4 is particularly interesting in that the presence of PX is obviously perturbing the massive Sedna planetoid and yet the average semi major axes of both remain nearly the same. The two massive objects move in synchrony, and in fact are essentially in one-to-one mean motion resonance.

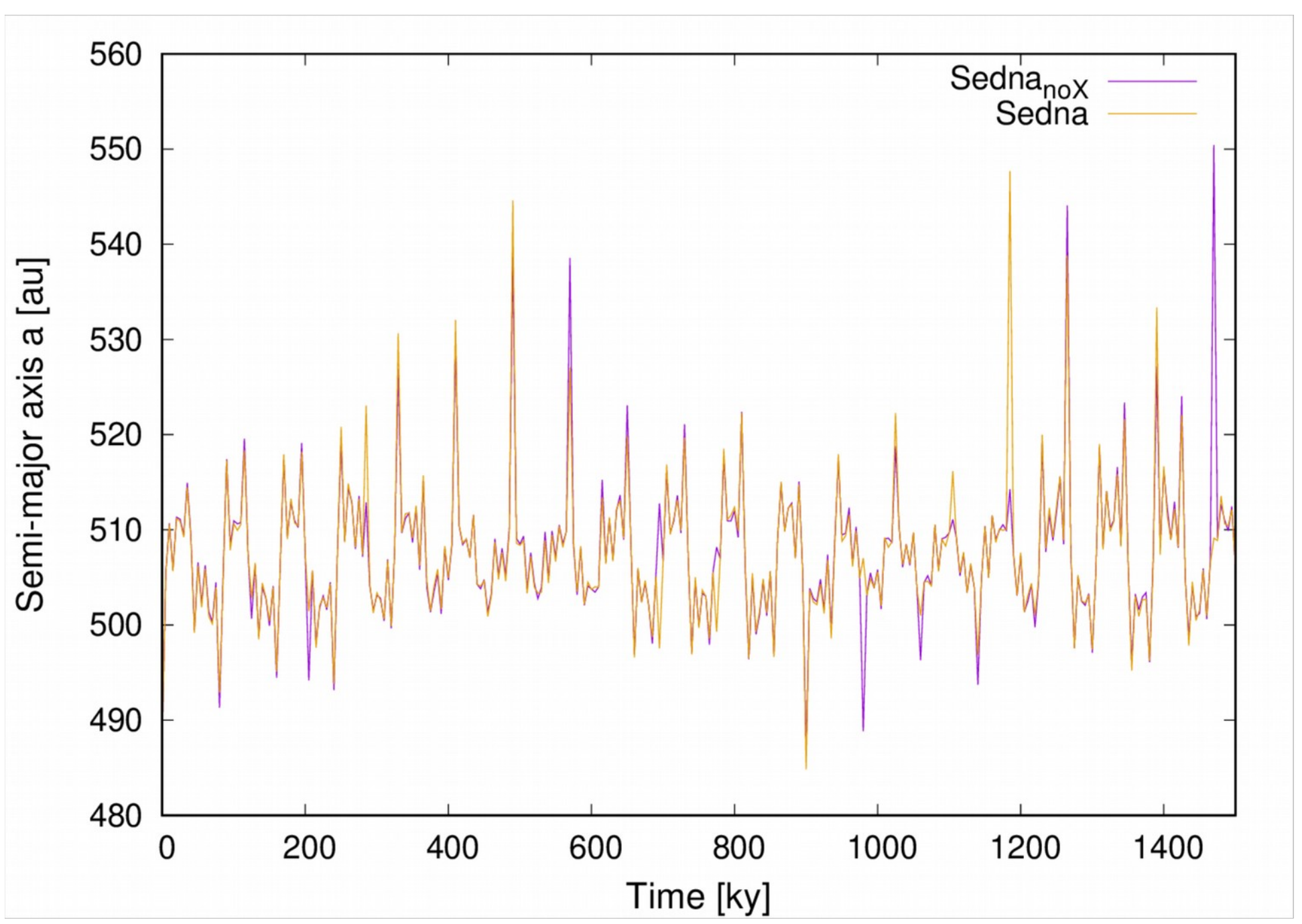

**Figure 3** Comparison of the semi-major axis of Sedna with and without PX

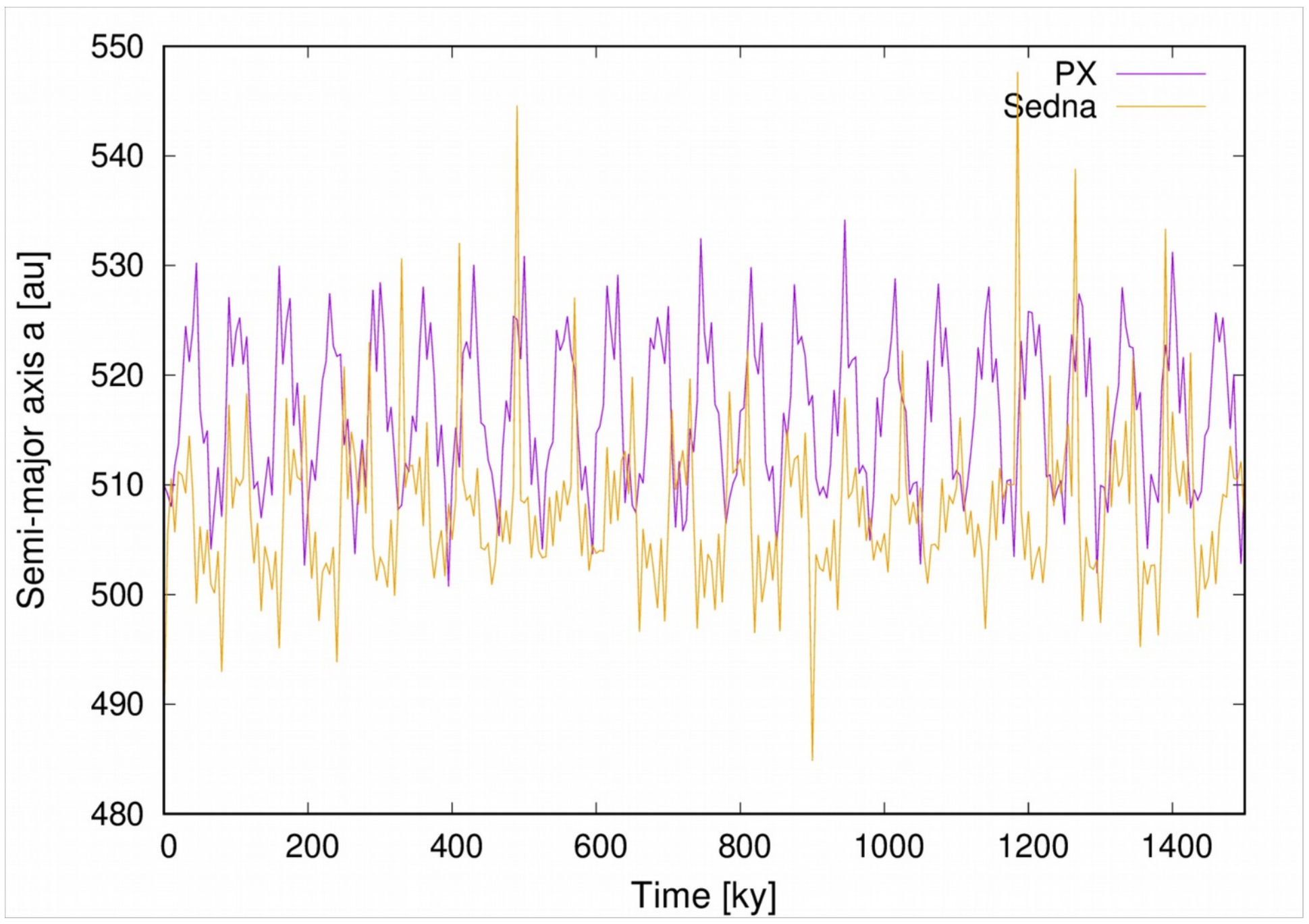

**Figure 4** Comparison of the semi-major axis of Sedna and PX.

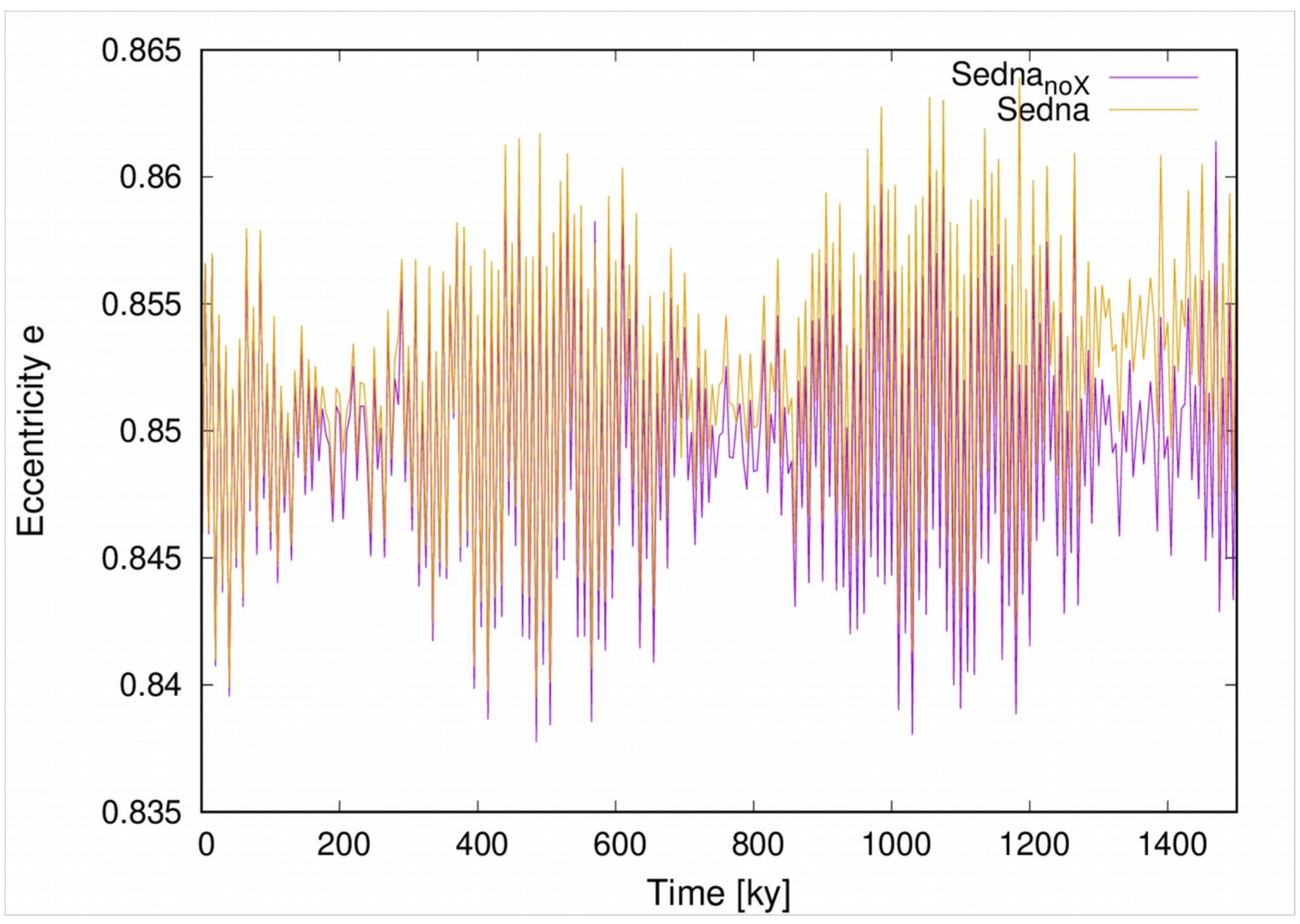

**Figure 5** Comparison of the eccentricity of Sedna with and without PX

Comparing the semi-major axis and the eccentricity for 1.5 Myrs we can see that the orbits of Sedna change minimally and are still stable (Fig. 3 and 4). Also its eccentricity (Fig. 5) does not change, but its eccentricity is higher than the average of the 12 TNOs used for this work because its orbits it is very close to PX in front of the other TNOs (Fig. 6).

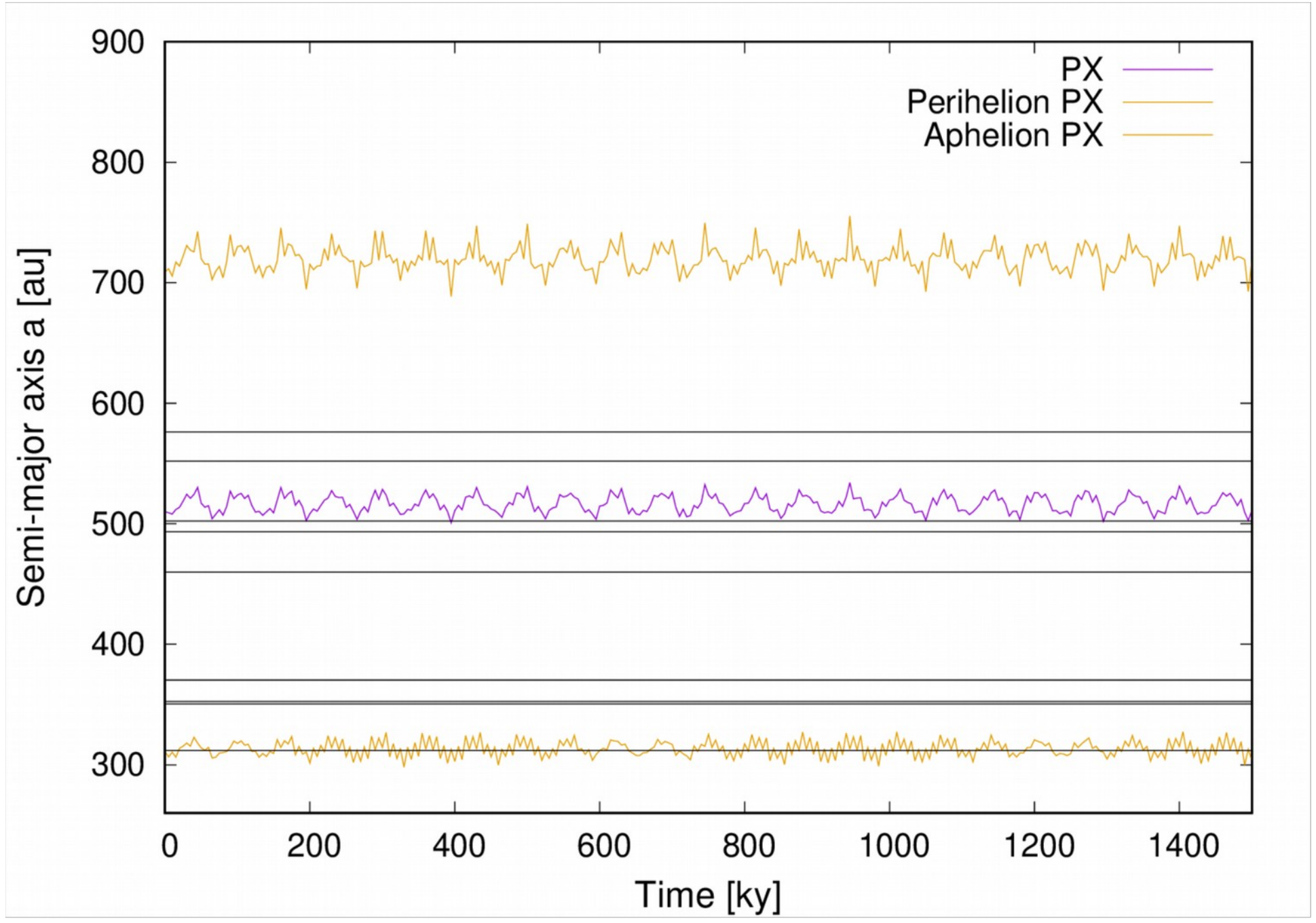

**Figure 6**. The space "cleaned" by PX: Perihelion, semi-major axis and Aphelion of PX. The lines describe the semi-major axis of the largest minor bodies inside the PX-region.

There are about 171 TNOs inside the PX orbit (semi major axes between 300 au and 700 au) and 9 large ones with absolute magnitude (H ≤ 7) inside this range (see Fig 7) with the relative osculating elements and absolute magnitude in Visual. All these bodies have large eccentricities (e > 0.8), their orbits might be being perturbed by PX, and most of them have inclinations similar to that of PX and greater 10 degree. The closest orbits to PX are the one of Sedna and in part of 2013 RA109 also. Very interesting is the situation with 2014 SR349 which lies completely within the PX's orbit including its perihelion, and this is unlikely to be a coincidence. Smaller bodies such as 2014 SR349 and 2015 FY561 and the comet C/1822 N1 (Pons) have near parabolic orbits, which indication past close encounters with PX. Regarding smaller bodies in the perihelion zone of PX are three comets C/1947 X1-A (Southern comet), C/1968 Y1 (Thomas) and C/2016 R2 (PANSTARRS), which might also be getting strongly perturbed by PX. Known minor bodies with semi-major axis similar to PX are almost all comets with very high inclination and near retrograde orbits, one of which in fact has a retrograde orbit (Comet Takamizawa).

| *Object Name* | *a (AU)* | *e* | *i (deg)* | *H* |
|---|---|---|---|---|
| 90377 Sedna (2003 VB12) | 551.9 | 0.8618 | 11.93 | 1.49 |
| 474640 Alicanto (2004 VN112) | 350.5 | 0.8649 | 25.49 | 6.46 |
| 523622 (2007 TG422) | 576.2 | 0.9382 | 18.57 | 6.47 |
| 765047 (2013 RA109) | 502.1 | 0.9082 | 12.39 | 6.21 |
| 768325 (2015 BP519) | 493.4 | 0.9284 | 54.09 | 4.34 |
| (2010 GB174) | 352.6 | 0.8627 | 21.55 | 6.74 |
| (2014 SR349) | 311.9 | 0.8480 | 17.95 | 6.46 |
| (2015 RX245) | 460 | 0.9009 | 12.13 | 6.20 |
| (2016 SD106) | 370.3 | 0.8847 | 4.80 | 6.82 |

**Figure 7**. Highly elliptical, long period TNOs with absolute magnitude H ≤ 7

## 4.0 Results and Discussion

The summary of the estimated orbital characteristics, mass and celestial location of Planet X, are included in the last two rows of Table 1 below, followed by a comparison chart in Fig. 8 of the size of Planet X with those of several planets and the TNO Sedna. This chart clearly shows the significant size of Planet X as double the size and gravitation of that of the Earth. This means a humans could walk on the planet with difficulty, which would be as if they were carrying a load equal to their own weight. Any possible residents would likely be very large and very strong.

The orbital elements, position and mass of PX were estimated using the methodology and models described in Section 2 above, the ephemeris data generator from the Jet Propulsion Lab (JPL) for the reference asteroids' locations, and Johnston's Archive (Johnston 2019) in sourcing the density data for approximating the TNO density values. The estimates of the planet's orbital characteristics and mass were calculated using two sets of data: the first 6 listed in the table from CalTech's original study and the second 6 selected by the authors using the criteria stated below. The ephemeris data for the asteroids and associated position estimates were based on the reference day of September 1, 2020. Since the mass values of the asteroids were unknown at the time, these were approximated using each asteroid's published diameter and the densities of similar sized asteroids from Johnston's Archive to arrive at three appropriate density ranges to approximate the mass of each. For reference asteroids with diameters greater than 900 km, the densities were estimated to be 1.96 g/cm$^3$; for diameters less than 900 km but greater than or equal to 400 km, the associated estimate was 1.06 g/cm$^3$; and for diameters less than 400 km the estimate was 0.94 g/cm$^3$. A published mass window is available for the dwarf planet Sedna, from which we used the more conservative, smaller value of the two limits at 1.70x10$^{21}$ kg. For the mass of the second largest body, the dwarf planet 2012 VP113, we used its published diameter together with the same density as that implied by the above mass value and diameter of Sedna, which yielded a mass approximation of 0.43x10$^{21}$ kg. In calculating an approximate diameter of Planet X, the density of the Earth was presumed to provide a lower bound, since Earth has the highest density of the known planets..

In calculating the diameter of Planet X the density was presumed to be equal to the density of the Earth. Selection of the asteroids was such that their longitude of perihelion falls within a 125° spread centered around 429° and 415° for the two sets of data, respectively. Inclinations below 35° and periods greater than 1800 years were also considered admissible. No other criteria were used in selecting the additional TNOs.

Although these selection criteria used were reasonably broad, the degree of synchronicity among the 12 bodies' orbits is striking. On calculating the average value of the asteroids' normal unit vectors, the orbital planes of all 12 were found to lie within about 13.5° of their average value with a maximum deviation of 20.0°. Further, on examining the longitude of perihelion of the asteroids, their azimuth orientation was found to lie in two distinct angular clusters, and again collective similarity of the orientations are surprising. The average offset of each cluster from the center line of the two clusters was estimated at 38.5° with an average deviation from this value of 7.6° using only the original 6 asteroids, and an offset of 36.0° and average deviation of 12.6° for all 12. Clearly, something is orchestrating this curious order and it would seem to be a large planet in a synergistic orbit with these 12 bodies, an order that was used in describing mathematical models to estimate a number of characteristics about this unseen planet.

Most astronomers claim not to know what causes this 'mysterious' organizing process, which is odd since in point of fact it was described by Romanian astronomers (Todoran & Roman 1992) over 30 years ago. The phenomenon is related to a planet's Lagrange points and the clustering of Trojan asteroids with planets having near circular orbits, of which there are several thousand in our own solar system orbiting in twin clusters around Jupiter, and even two at current count orbiting the Earth. In this situation the planet's two off axis Lagrange points L4 and L5 pull asteroids into two clusters, which move ahead and behind the planet in its orbit, clustered around the rotating Lagrange points. For planets in prominently elliptical orbits the process is more complex. In this situation the Lagrange points come closer together near the Sun as the planet moves towards perihelion, always remaining the same distance to the planet as to the Sun. In response the two clusters of asteroids move in highly elliptical orbits, whose major axes form a spatially stable, V- shaped pattern, which both sets of data clearly indicate. When the planet gets close enough to perihelion, the associated off axis Lagrange point disappears, since the distance symmetry from the planet to the Sun and to the Lagrange's 'desired' position no longer exists geometrically. Given this scenario we expect the planet to be orbiting in the same plane as those of the asteroids, and to have its major axis approximately midway between the two clusters, bisecting the V. With this presumption the planet's inclination and longitude of the ascending node were readily calculated using the planet's estimated normal unit vector, taken as the normalized average of the asteroid's associated unit vectors in each data set. Similarly, using the longitude of perihelion of the asteroids and their distribution in the two clusters, the azimuth orientation of Planet X's major axis was estimated to be that angle which bisects the V formation, allowing estimates of its longitude of perihelion of about 428.6° for the original 6 asteroids and 415.1° for all 12. From these averages we were able to estimate the planet's argument of perihelion directly, and therefore having estimates of all three of the planet's orbital coordinates, we were able to construct its complete transformation matrix $\boldsymbol{U_p}$ which as stated includes its azimuth and normal unit vectors as two rows.

With the angular width of the asteroids' V pattern, which turns out to be about 72°, the shape of the planet's orbit has to be 'wide' enough to encompass a substantial portion of both sides of the V, which means its eccentricity has to be small enough to permit this expanded orbital width. For this to happen, and with the planet's off axis Lagrange points having their largest influence in the upper part of its orbit near aphelion, the azimuth orientation in space where both planet and asteroids spend most of their time is very close to the same on average and also fairly close to the location of the off axis Lagrange points when the planet is not too far from aphelion. That said, we presumed that the average range over time of the planet should have approximately the same angular orientation at that point as average value of those of the reference asteroids. The true anomalies are quite a bit different, however, due to the fact that the major axes of the asteroids and planet differ by about 36° as discussed above. With this attribute a simple expression (20) was developed for estimating the planet's eccentricity in terms of the asteroids' true anomalies at the location of their average ranges averaged over time. Using this methodology the estimates produced for the eccentricity were 0.37 for the case of the original 6 asteroids and 0.39 using all 12.

The length of Planet X's semi major axis was also found to depend on a significantly less eccentric orbit for the planet. Given that Planet X is heavily influencing the asteroids' orbits, this effect also includes the dynamics of their collective centripetal acceleration. When we examined each asteroid's smallest value of centripetal acceleration, which occurs at the two end points of their minor axes, there is another more subtle pattern: the shorter the length of its semi major axis and / or the smaller its eccentricity, the larger its centripetal acceleration at that point. This means in effect that we would expect the centripetal acceleration to be at neither extreme of the asteroids' values, but rather somewhere closer to their center, since Planet X's motion is clearly influencing the asteroids motion in a wide range of locations as well. Accordingly, we used the planet's centripetal acceleration at this key point on its orbit as the average of those of the reference asteroids at the same point, from which the semi major axis length could be readily calculated, from which we estimated 519 AU for the 6 asteroid case and 450 AU for the case with 12. The much higher number is due to the fact that the smaller date set severely under represents shorter period asteroids, and so discounts them in the computation. The original data set has no asteroids at all with a semi major axis length less than 200 AU, whereas the expanded data set contains three. In either case, however, the axis length is not verry far from 500 AU, corresponding to an orbital period of 11180 years, and this this value happens to be close to the planet's actual period, then four of the reference asteroids are in near mean motion resonance with it (and arguably a fifth a well, but this one was omitted from the analysis): Sedna, 2000 OO67, 2013RF98, and 2004 VN112, the first two in a 1:1 resonance the last two in 2:1. Presuming these resonances to exist, we used their average long resonance period of 11512 years to estimate a more precise value of the semi major axis length using Kepler's third law, which turned out to be 510 AU. This number should be very close to reality given that a third of the reference asteroids are very likely in orbital resonance with the planet's orbit.

One of the most interesting questions about this study all along has been about how big can this rogue planet be. Once again the asteroids point us to an interestinge approximation. The method used in estimating the mass involves a corollary to the law of angular moment, which states that if we take the mass of each body in a gravitational system, multiply each body's mass by its vector distance from the system's center of mass and sum up all the resulting mass-weighted range vectors, the result remains zero over time. This is really no great surprise since this statement is virtually equivalent to the definition of the system's center of mass. Using this fact and knowing the orbits and estimated mass values of the reference asteroids orbits, a second order quadratic expression for the planets mass was derived in terms of the mass-weighted range vectors of the asteroids and the Sun, the planet's mass being that value needed to maintain the overall symmetry. On taking expected values of each side using the range vectors averaged over their true anomalies, the second order estimate of the planet's mass follows immediately. From this expression the estimates calculated were 7.4 Earth masses using the original 6 asteroids and 7.1 using all 12. It should be noted that the sum of the mass-weighted range vectors of the reference asteroids is located very near the systems' center of gravity because their cumulative mass is at least four orders of magnitude smaller than that of Planet X, a useful fact that aided the mass estimation and planet location development approaches. An obvious caveat to this approach is that as more asteroids are found that are also under the planet's similar clustering influence, although the mass estimate will increase slowly as one half the newly discovered mass divided by the existing reference asteroids' total mass.

To estimate the planet's recent location we used a slightly altered version of equation (33). The range vector of each asteroid was computed using the specific celestial locations of the reference asteroids provided and the ephemeris data, epoch 2000, available at the Jet Propulsion Laboratory's web site for each for the reference asteroids associated with the reference day of September 1, 2020, together with the planet's estimated mass for the case using 12 asteroids. The formalism for the estimation is described in Section 2.5 tells us that given a statistically comprehensive data set, the sum of asteroids' mass-weighted range vectors effectively points to the planet's celestial location. But neither of the two data sets are in any sense "comprehensive", so we have an issue. Our data was found to have a strong bias which is currently unavoidable: the long period asteroids available for the estimate are known to astronomy because each is near its perihelion and therefore close enough to the Earth to be observed. It would obviously be preferable to have a large number of asteroids scattered randomly about their orbits, but this simply is not now the case, and all researchers have the restriction of what astronomy's current state of knowledge is regarding long period TNOs, and it is limited. And yet this said, there still is a credible way to proceed, although cautiously.

**Table 1** Reference asteroid orbital data and estimates of Planet X's orbital parameters, mass and location on the day Sept 1, 2020 (epoch 2456100.5)

| TNO | Period ( yrs ) | *a* (AU) | e | Perihelion ( AU ) | ι ( Deg ) | Ω ( Deg ) | ω ( Deg ) | Diameter ( km ) | Mass ( kg ) | ϖ ( Deg ) | α (h:m:s / Deg) | δ (d:m:s / Deg) | r (AU) |
|---|---|---|---|---|---|---|---|---|---|---|---|---|---|
| 2012 VP113 | 4473 | 272 | 0.70 | 81 | 24.0 | 90.8 | 293.8 | 702 | 4.300 e20 | 384.6 | 03 41 50.28<br>55.4583 | +04 02 15.2<br>+4.0375 | 83.80 |
| 2 013 RF98 | 5600 | 349 | 0.88 | 36 | 29.6 | 67.6 | 316.5 | 55 | 8.210e16 | 384.1 | 03 14 30.00<br>48.6250 | +05 51 11.0<br>+05.8531 | 37.29 |
| 2004 VN112 | 5850 | 315 | 0.85 | 47 | 25.6 | 66.0 | 327.0 | 152 | 1.728 e18 | 393.0 | 03 26 17.08<br>51.5708 | +11 24 40.9<br>+11.4114 | 47.93 |
| 2007 TG422 | 12789 | 472 | 0.93 | 36 | 18.6 | 113.0 | 285.8 | 174 | 5.384 e18 | 398.8 | 04 59 36.86<br>74.9042 | +10 34 13.1<br>+10.5703 | 38.69 |
| 90377 Sedna | 11390 | 506 | 0.85 | 76 | 11.9 | 144.5 | 311.5 | 995 | 1.700 e21 | 456.0 | 03 55 03.92<br>58.7667 | +08 08 15.7<br>+8.1378 | 84.26 |
| 2010 GB174 | 6945 | 349 | 0.87 | 48 | 21.5 | 130.7 | 347.7 | 250 | 7.688 e18 | 478.4 | 13 06 13.30<br>196.5542 | +13 09 48.1<br>+13.1633 | 74.04 |
| 2000 CR105 | 3442 | 228 | 0.81 | 44 | 22.7 | 128.3 | 317.0 | 285 | 1.127 e19 | 445.3 | 10 39 04.58<br>159.7708 | +19 26 53.8<br>+19.4483 | 63.88 |
| 2014 SR349 | 4960 | 299 | 0.84 | 48 | 18.0 | 34.8 | 341.5 | 200 | 3.936 e18 | 376.3 | 22 44 20.43<br>341.0833 | -25 36 41.1<br>-25.6114 | 53.75 |
| 2010 VZ98 | 1846 | 150 | 0.77 | 34 | 4.5 | 117.4 | 313.9 | 401 | 3.550 e19 | 431.3 | 03 38 15.28<br>54.8125 | +15 17 26.9<br>+15.2908 | 34.77 |
| 2003 SS422 | 2744 | 196 | 0.80 | 39 | 16.8 | 151.1 | 210.8 | 168 | 2.333 e18 | 361.9 | 01 43 12.65<br>25.8000 | -03 37 23.7<br>-03.6233 | 39.71 |
| 2000 OO67 | 11760 | 531 | 0.96 | 21 | 20.1 | 142.3 | 212.2 | 44 | 4.191 e16 | 354.5 | 04 15 23.07<br>63.8458 | +00 56 14.5<br>+0.9372 | 28.76 |
| 2013 UT15 | 2740 | 196 | 0.78 | 44 | 10.7 | 191.9 | 252.3 | 274 | 1.012 e19 | 444.2 | 01 22 30.0<br>20.6250 | +06 50 42.0<br>+6.8450 | 57.00 |
| Planet X (6 TNOs) | 11430 | 507 | 0.37 | 319 | 19.45 | 95.05 | 333.6 | 24600 | 4.440 e25 | 428.7 | N/A | N/A | N/A |
| Planet X (12TNOs) | 11512 | 510 | 0.39 | 311 | 14.64 | 107.76 | 307.3 | 24000 | 4.230 e25 | 415.1 | 03 52 41.0<br>+58.17 ° | +7° 22 48.0<br>+7.38 ° | N/A |

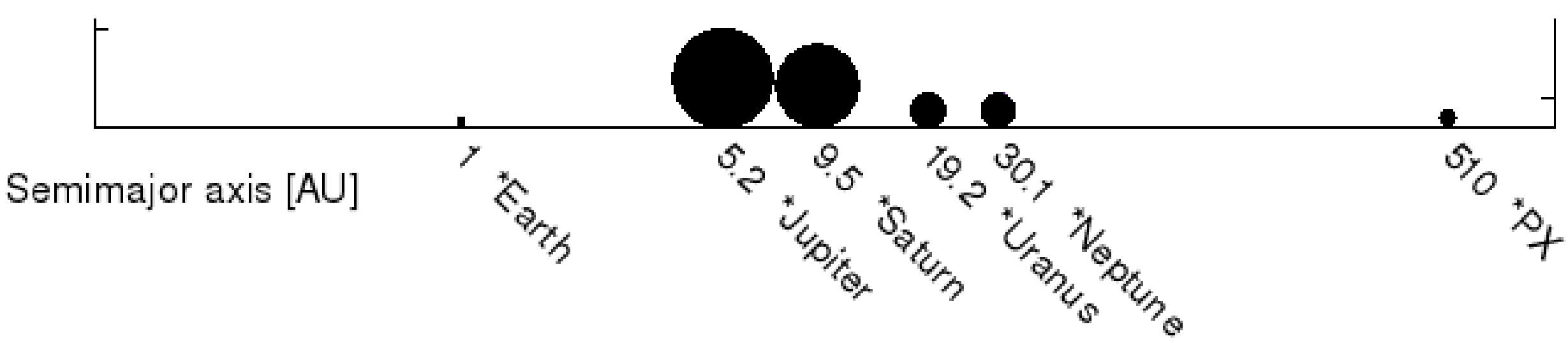

**Figure 8.** Size distribution of the giant planets with planet Earth, PX and Sedna. The Circles are the diameter of each body in scale with Jupiter

The two most massive asteroids of our data set, Sedna and 2012 VP113, are in opposite clusters and are both close to perihelion and in fact fairly close to each other as well; this fact enables a solution. These two bodies have more mass than the other 10 asteroids combined, but with that mass still being at least four orders of magnitude smaller than that of Planet X, their mass-weighted range vectors have to be very close to the system's center of mass. Given this fact, then Planet X's location should also in theory be very close to its perihelion as well since the center of mass has to be between Planet X and the Sun, so the location formalism should give a reasonable estimate of its location on the reference day in 2020. That estimate gives an approximate value of right ascension 3h 52m 41.0s and declination +7° 22' 48.0", placing the planet's celestial coordinates in the lower region of Taurus on the reference day of September 1, 2020. Using the methodology in Section 2.6,, the planet's perihelion location was computed, which depends on the asteroids' orbital parameters but not their instantaneous celestial coordinates. The results obtained for the right ascension was 3h 45m 9.5s and for the declination, +8° 3' 36", also in the lower region of Taurus. Similarly, the aphelion's location was located at right ascension 15h 45m 9.5s and declination – 8° 3' 36" in the upper region of Libra. Although some of the recent searches have been in Taurus, even if the planet is near perihelion that still means its being over 300 AU from the Earth nevertheless, no small challenge to locate. The asteroids' approximate apparent airless visual magnitude and surface brightness using the standard IAU H-G system magnitude model (see also Horizon JPL or our previous paper (6) ), using a Phase angle equal to 0.68 should have a value of APmag $\approx$ 26 mag. For comparison this is about 2 mag fainter than the TNO 505478 (2013 UT15), observing it with the same phase angle, meaning that such a body might be seen by an 8m-telescope (since its magnitude might reach 27 mag) or a space telescope such as Hubble, but almost impossible with telescopes with a diameter smaller than 8m.

## 4. Conclusion

The approach used for deducing the existence of a hypothetical Planet X has been to create a set of mathematical models that relate the orbital characteristics and interrelationships of 12 reference asteroids to those of a hypothetical though increasingly likely large planet, consistent with the asteroids' known characteristics and behavior. The reference asteroids used in the analysis were the 6 from the CalTech astronomers' original study together with an additional 6 selected by the authors using similar but significantly broader selection criteria: TNOs with longitudes of perihelion within a 350° to 480° range, inclinations below 35° and periods exceeding 1800 years were all admissible with no constraints on mass and eccentricity. JPL's ephemeris generator was used for determining asteroid locations on the reference day of September 1, 2020, Johnston's Archives for representative TNO density profiles, and widely published data for the orbital data of the 12 TNOs.

On calculating the unit vector normal to the plane of each asteroid, it was found that each of the 12 reference asteroids' orientations lies close to the same plane, and in fact they deviate from the average plane by only 13.5° on average. That orientation was presumed to be the plane of Planet X, which enabled calculation of its inclination at +14.6 ° and longitude of the ascending node at 107.7 °, uniquely determining the plane's spatial orientation. One of the more striking aspects of the asteroids' orbits is their azimuth orientations, as seen by their longitudes of perihelion, which were found to fall into two distinct azimuth clusters, centered around a location in the upper region of Libra. The two clusters were found to be about 72° apart on average with azimuth angles deviating from the average center line of their respective cluster by an average of 12.5°. The asteroids' major axes therefore were found to form a distinct V pattern, the center of which is most likely according to theory to be a close approximation to the azimuth orientation of the planet's major axis, which was presumed. This clustering phenomena was studied by Romanian astronomers in the early 90s, and was found to arise with large planets in elliptical orbits organizing the orbits of asteroids using in part the planet's two off axis Lagrange points L4 and L5, which move along the planet's orbit as it approaches and exits perihelion, forming the stable V pattern of the asteroids' major axes over time, which seems to be the case with the12 reference asteroids studied. With the azimuth center line and the planet's orbital plane estimated, this result was found to uniquely determine the planet's argument of perihelion at 307.3 °.

The planet's eccentricity was derived from those of the asteroids at their expected range values averaged over time, and was found to be considerably smaller than those of the asteroids': 0.37 for the case of the 6 asteroids and 0.39 for the case with all 12. A large portion of the planet's orbit nearest perihelion completely 'surrounds' over half of the asteroids' orbital V pattern nearest perihelion, due to its much wider minor axis than those of the asteroids long semi major axis length.

The semi major axis length was roughly approximated using the average of the asteroids' centripetal accelerations at their intersections with their minor axes, which gave estimates of 519 AU for the data set of 6 TNOs and to 450 for the data set of 12. From these estimates, it was found that the corresponding orbital period computed from the approximate length is fairly close to being in mean motion resonance with four and possibly five of the reference asteroids! We took this fact as an indication that our rough estimate was most likely approximating a slightly different major axis length, an indication that we used to estimate the orbital period that best agreed with the four most likely resonance candidates, which produced a period of 11512 years and corresponding semi major axis length of 510 AU, calculated from Kepler's third law. Supposing that Planet X is now near its perihelion as we predict, it is interesting that its last perihelion corresponds to the Younger Dryas event for the Earth, about 12,000 years ago, during which Earth experienced a significant event, a period of abrupt climate change that interrupted the warming trend after the last Ice Age.

The planet's mass was computed using the known symmetry of the mass-weighted position vectors associated with bodies orbiting a massive central body with a large gravitational field, in our case of course the Sun. From this principle we know that the sum of the associated mass-weighted range vectors of the Sun, planet and TNOs, each measured from the system's center of mass, remains zero over time, which allows the planet's mass to be computed by determining its magnitude needed to maintain the balance. The resulting equation describes the instantaneous condition of the relationship, so to solve for the planet's mass we took the expected value of both sides, which resulted in the final quadratic expression. For the data set of 6 TNO's the planet's mass estimate was found to be 7.4 Earth masses, and 7.1 masses for the set of 12.

Estimating the recent location of Planet X was greatly aided by the two most massive asteroids, Sedna and 2012 VP 113, by their being in different clusters, close proximity to perihelion and close to each other on the reference day in 2020. So even though the data sets are small and all the asteroids are biased in that all the 'organized' asteroids available are near perihelion, these facts made the estimation formalism considerably more reasonable. The asteroids' mass-weighted ranges have to be close to the center of mass which is always between the Sun and Planet X, which apparently is the case, given that the sum of all the mass-weighted ranges have to sum to zero. That said, our estimates predict that the planet's location and perihelion were both estimated to be in the lower region of Taurus on the reference day and the aphelion to be in the upper region of Libra. Interestingly, in April 2025 a paper (Phan, Goto, et al. 2025) came out on ArXiv by nine scientists from several researcher in Taiwan, Japan and Australia, who collaborated to analyze IRAS and AKARI satellite survey data over a 23 year period, from which they were able to isolate a dwarf planet (designated 2017OF201) in the data that could possibly be under the influenced of Planet X and in fact may be in a 2:1 mean motion resonance with Planet X as well. The object was in Taurus and they estimated it to have a heliocentric distance range of 500 to 700 AU and a mass of 7 to 17 Earth masses. Astronomers the world over will be closely following their work.

Finally we computed an apparent visual magnitude, which we conclude that due to our result (PX has an Apmag around 26 mag), the only way to observe it is using a very large telescope with a diameter at least larger than 8m and/or a space telescope such as e.g. Hubble or JamesWebb. PX's magnitude might be very similar to that of Kepler-10c or 55 Cancri. Interesting also is its gravitational acceleration, which would be 19.6 $m/s^2$., which means a human could move slowly there and with difficulty. Supposing theoretically it is inhabited, its inhabitants would be much larger and stronger than a human-being and very likely very sensitive to light. Here is an artistic representation of PX through its day-light and night shadow (see Fig. 9 below).

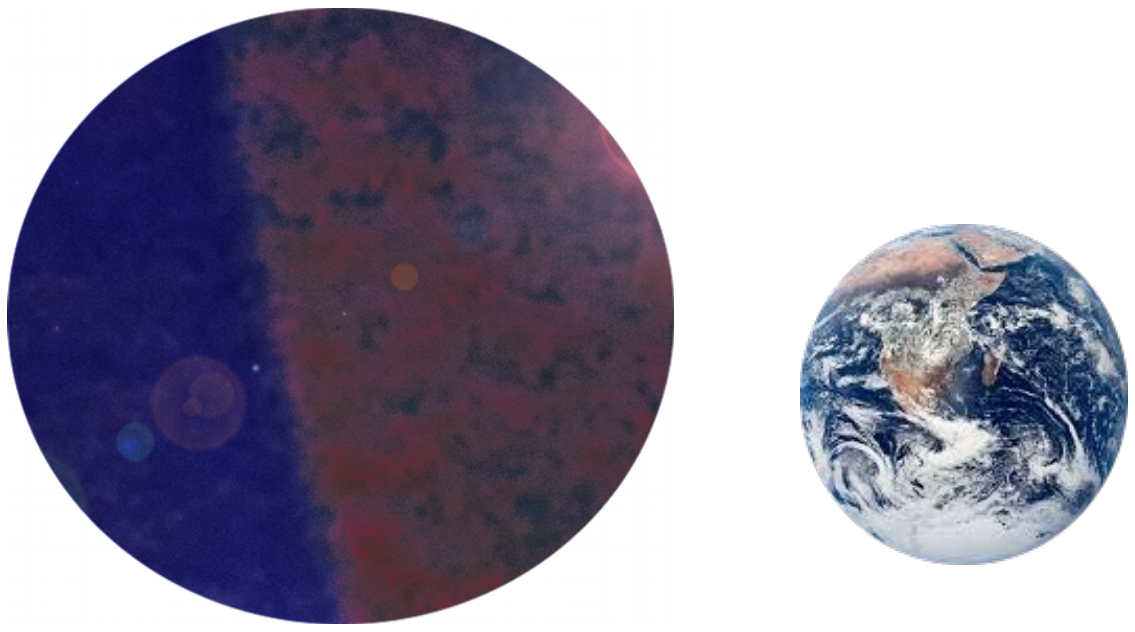

**Figure 9** Artistic representation of Planet X with its thick and dense atmosphere. The Sun-light is on the right-top. The Earth is in comparison for the size.

Many TNOs with semi major axes between 300 au and 700 au may have their orbital evolution influenced by PX, in fact the majority of these bodies have very large eccentricities and some nearly in retrograde orbits, with only 1 comet in retrograde orbit with very similar semi-major axis to that of PX. One body has the semi-major axis at the PX perihelion, 2014 SR349 and almost all the bodies with similar semi-major axis similar to PX show cometary activity. We have a known cluster of 171 TNOs in the PX region, the majority of them are far Centaurs.

The stunning coordination among 12 objects in long period orbits is certainly no coincidence. Something very large is causing it and once observed that something will change our idea of the solar system forever. We hope that this paper may help the process of discovery by adding to the analytical tools available and to the body of evidence for Planet X's existence published to date.

Our simulations have shown that the orbit of Sedna is stable with that of PX's orbit predicted by our analysis. And so we close with an intriguing question: if our predictions are valid, might Sedna be actually a co-orbital object of Planet X?